\documentclass{elsart}

\usepackage{amssymb}
\usepackage{epsf,colordvi,color,amsbsy}
\usepackage[dvips]{graphicx}
\begin{document}

\begin{frontmatter}

\title{Wavelet analysis of solar magnetic strength indices}

\author[]{Stefano Sello \corauthref{}}

\corauth[]{stefano.sello@enel.it}

\address{Mathematical and Physical Models, Enel Research, Pisa - Italy}

\begin{abstract}

Wavelet analysis of different solar activity indices, sunspot numbers, sunspot areas, flare index, magnetic fields, etc., allows us to investigate the time evolution of some specific features of the solar activity and the underlying dynamo mechanism. The main aim of this work is the analysis of the time-frequency behavior of some magnetic strengtht indices currently taken at the Mt. Wilson Observatory 150-Foot Solar Tower. In particular, we analyzed both the long time series (Jan 19, 1970 - Jan 22, 2013) of the Magnetic Plage Strength Index (MPSI) values and of the Mt. Wilson Sunspot Index (MWSI) values, covering the descending phase of cycle 20, the full solar cycles 21-23 and the current part of the 24 solar cycle. This study is a further contribution to detect the changes in the multiscale quasiperiodic variations in the integrated magnetic solar activity with a comparison between past solar cycles and the current one, which is one of the weaker recorded in the past 100 years. Indeed, it is well known that an unusual and deep solar minimum occurred between solar cycles 23 and 24 and the time-frequency behavior of some magnetic strengtht indices can help to better interpret the responsible mechanisms. 

\end{abstract}
\end{frontmatter}

\section{Introduction}

Observations of solar magnetic fields both in plage areas and in sunspot penumbrae and umbrae have been sistematically carried out by different Solar Observatories such as the National Solar Observatory's McMath-Pierce atop Kitt Peak and the 150-Foot Solar Tower at Mt. Wilson. For example, M.J. Penn and W. Livingston, 2006, 2010, recently observed a decrease in the sunspot magnetic field strength using the Zeeman-split 1564.8 nm Fe I spectral line at the NSO Kitt Peak McMath-Pierce telescope and this trend was seen to continue in observations of the first sunspots of the new solar Cycle 24. Moreover, extrapolating a linear fit to this trend, would lead to only half the number of sunspots during Cycle 24, compared to Cycle 23, and this imply virtually no sunspots at all in Cycle 25.
On the other hand, for each magnetogram taken at the 150-Foot Solar Tower at Mt. Wilson Observatory operated by UCLA, a Magnetic Plage Strength Index (MPSI) value and a Mt. Wilson Sunspot Index (MWSI) value are continuously calculated since Jan 19 1970. To determine MPSI the authors sum the absolute values of the magnetic field strengths for all pixels of the solar disk where the absolute value of the magnetic field strength is between 10 and 100 gauss. This number is then divided by the total of number of pixels (regardless of magnetic field strength) in the magnetogram. The MWSI values are determined in much the same manner as the MPSI, but here the summation is only done for pixels where the absolute value of the magnetic field strength is greater than 100 gauss. In this work we performed a wavelet analysis on magnetic field strength indices using the released text file containing the daily average of MPSI and MWSI values. The goal is to give a further contribution to better follow the current decreasing epoch of the solar activity as well evidenced by the long-term behavior of the solar magnetic field strength.
As many previous works showed, one method to shed more light into underlying basic processes in the solar interior and dynamo mechanism that govern the nature and evolution of solar photospheric magnetic fields, is by studying different scale periodicities generated by various surface activity features at different times in different solar cycles. Many significant quasi-periodicities (both local and persistent in time) have been detected using both Fourier and wavelet analyses with different solar activity indices. As an example, the well known and discussed near 158 days(d) Rieger periodicity is reported in the solar flare occurrence rates (Oliver, Ballester, and Baudin, 1998) and suggested to be correlated to the periodic emergence of magnetic flux through the photosphere. Recently, A.K. Bisoi et al., 2013 performed a detailed analysis of the NSO/KP synoptic magnetic database during the period 1975.14 - 2009.86 using both wavelet and Fourier techniques in order to understand the role of periodic changes, if any, in the solar photospheric fields leading to the recent deep minimum at the end of cycle 23.

In the present work we analyzed, using a proper wavelet methodology, the two time series of the Magnetic Plage Strength Index (MPSI) value and the Mt. Wilson Sunspot Index (MWSI) in the time interval: Jan 19 1970 - Jan 22 2013. These indices are an integral measure (spatially unresolved) of the magnetic field strength covering only two magnetic bands of intensity: a low intensity with $10-100$ gauss and a high intensity with $> 100$ gauss, respectively.  

\section{Wavelet analysis}

Since the introduction of the wavelet transform by Grossmann and Morlet, 1984, in order to overcome the window limitations of the Gabor transform to non-stationary signals, this technique has been extensively applied in time-series analysis, including the study of solar and stellar activity cycles (Ochadlick,et al. 1993, Lawrence, et al. 1995, Oliver, et al. 1998, Sello 2000, 2003). With a local decomposition of a multiscale signal, wavelet  analysis is able to properly detect time evolutions of the frequency distribution. This is particularly important when we consider intermittent and, more generally, non-stationary processes. More precisely, the continuous wavelet transform represents on optimal localized decomposition of a real, finite energy, time series: $x(t) \in L{^2}(\Re)$ as a function of both time, $t$, and frequency (scale), $a$, from a convolution integral:

\begin{equation}
(W_{\psi}x)(a,\tau) = \frac{1}{\sqrt{a}} \int_{-\infty}^{+\infty}
dt~x(t)\psi^* (\frac {t-\tau}{a})
\label{eq:wav1}
\end{equation}

where $\psi$ is called analysing wavelet if it verifies an admissibility condition:

\begin{equation}
c_{\psi} =  \int_{0}^{+\infty}
d \omega ~ \omega^{-1} {\arrowvert \hat{\psi}(\omega) \arrowvert}^2 < \infty
\label{eq:adm1}
\end{equation}

with:

\begin{equation}
\hat{\psi}(\omega)=\int_{-\infty}^{+\infty}
dt~ \psi(t) \mathrm{e}^{-i \omega t}
\label{eq:adm2}
\end{equation}

This last condition imposes: $\hat{\psi}(0)=0$, i.e. the wavelet has a zero mean. In eq.(\ref{eq:wav1}) $a,\tau \in {\Re}, (a \neq 0)$ are the scale and translation parameters, respectively (Daubechies, 1990, Mallat, 1998).
In fact, it follows from eq.(\ref{eq:wav1}) that the effectiveness of the wavelet analysis depends 
on a suitable choice of the analyzing wavelet for the signal of interest. For our time-series application, where we are mainly interested to track the temporal evolution 
of both the amplitude and phase of solar activity signals, we chosed to use the family of complex analyzing wavelets
consisting of a plane wave modulated by a Gaussian, called Morlet wavelet (Torrence and Compo, 1998):

\begin{equation}
\psi(\eta) =  \pi ^{-1/4} \mathrm{e}^{i\omega_{0} \eta} \mathrm{e}^{-\eta^{2}/{2\sigma^2}}
\label{eq:wav4}
\end{equation}

where: $\eta={{t- \tau} \over {a}}$, and $\omega_{0}$ is a non-dimensional frequency. $\sigma$ is an adjustable parameter which can be determined in order to obtain the optimal wavelet resolution level both in time and frequency, for the characteristic time-scale of the original series (Soon, et al. 1999).
The limited frequency resolution imposes an half-power bandwidth of our wavelet given by: ${\Delta f \over f} \approx 0.12$.
The local wavelet spectrum at frequency $k$ and time $t$ and generally visualized by proper color contour maps is:
 
\begin{equation}
P(k,t) = {1 \over {2c_\psi k_0}} {\arrowvert W({k_0 \over k},t) \arrowvert} ^2,~k \geq 0
\label{eq:ent3}
\end{equation}

In eq.(\ref{eq:ent3}) $k_0$ is the peak frequency of the analyzing wavelet $\psi$ (Torrence and Compo, 1998). 

\section{Data selection}

The magnetic solar activity indices used for our analysis are those released by the 150-Foot Solar Tower at Mt. Wilson Observatory: 1) a Magnetic Plage Strength Index (MPSI) value and 2) a Mt. Wilson Sunspot Index (MWSI) value. We recall here that to determine MPSI the authors sum the absolute values of the magnetic field strengths for all pixels where the absolute value of the magnetic field strength is between 10 and 100 gauss. This number is then divided by the total of number of pixels (regardless of magnetic field strength) in the magnetogram. The MWSI values are determined in much the same manner as the MPSI, but here the summation is only done for pixels where the absolute value of the magnetic field strength is greater than 100 gauss. The current text file containing the daily average MPSI and MWSI values considers data in the interval: Jan 19, 1970 - Jan 22, 2013 with 11353 data for each index. The time value is in julian date where: May 23, 1968 at noon U.T. = julian day 2440000.0 = j0.0. The above series are not evenly spaced and the mean time step is: 1.38 days.

\section{Results and discussion}

The principal information derived from the application of the wavelet formalism to real data is the local wavelet spectrum, eq.(\ref{eq:ent3}), which allows us to resolve the time-evolution of the related frequency distribution.
In Fig. 1 we show the results of the wavelet analysis applied to the first magnetic index: MPSI. The time interval is: Jan 19, 1970 - Jan 22, 2013 or in julian date:  2440606.24 - 2456315.37 covering the solar cycles: 20 (descending phase), 21-23 and current 24. The upper panel shows the original time series in its natural units. The central panel shows the amplitudes of the wavelet local power spectrum in terms of a color contour map. Horizontal time axis corresponds to the axis of the upper time series and the vertical scale (log frequency values in red) axis is expressed in period (days d or years yr). Here the range analyzed is: 2.8 d - 40 yr.
The right panel shows the global wavelet spectrum (an averaged and weighted Fourier spectrum) obtained by a time integration of the local map for each frequency. The significance of the local power map was tested using an adjustable red-noise autoregressive lag-1 background spectrum (Torrence and Compo, 1998).

\begin{figure}[h!]
\resizebox{\hsize}{!}{\includegraphics{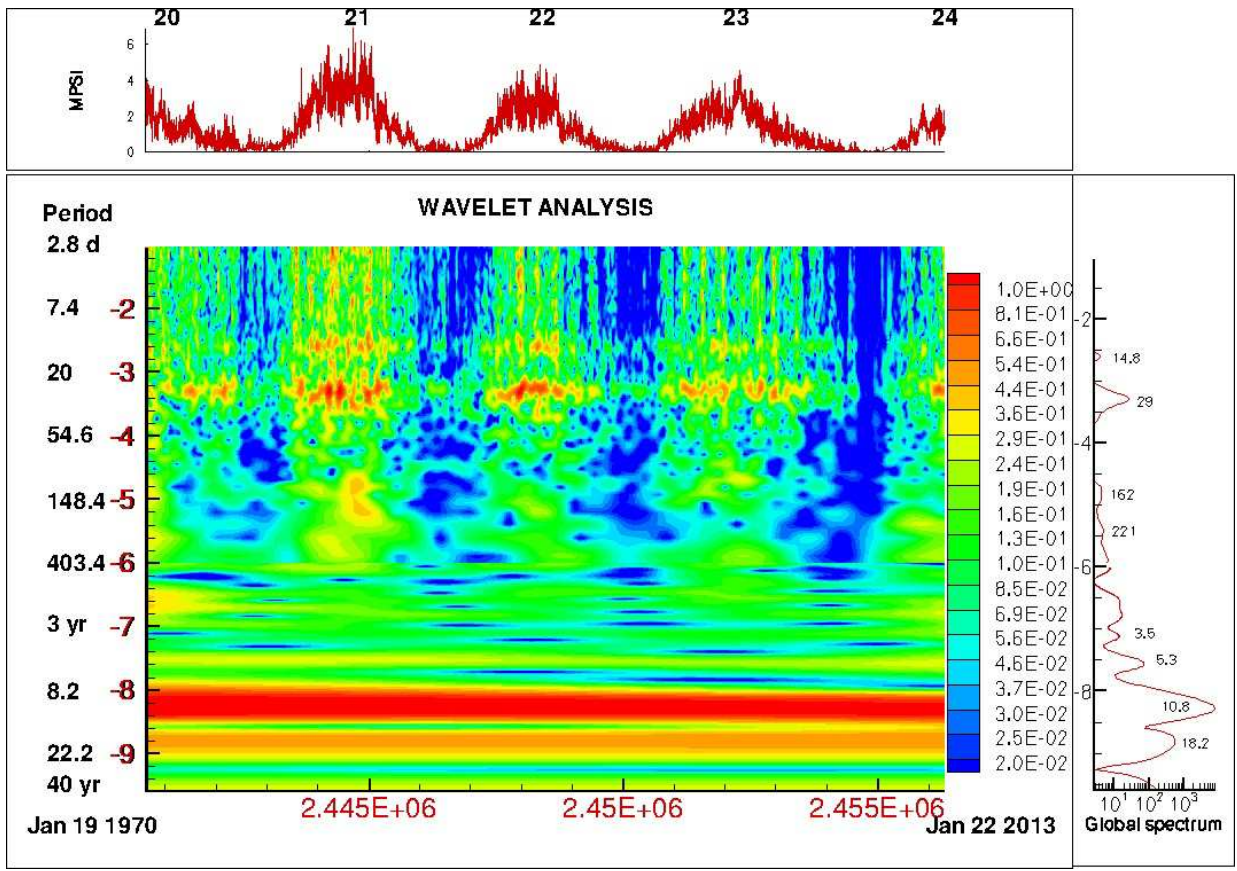}}
 \caption{Wavelet analysis of magnetic MPSI index. See text for details.}
 \label{fig1}
\end{figure}

The main periodicities detected are in decreasing order:

1) 18.2 yr 2) 10.8 yr; 3) 5.3 yr; 4) 3.5 yr
5) 221 d; 6) 162 d; 7) 29 d; 8) 14.8 d

as we can see in the global power spectrum.
The strongest periodicities are, of course, the classic Schwabe period of near 11 yr and the rotational periodicity near 29 d with related subharmonics. It is interesting to note the variation in intensity and distribution of these periodicities from solar cycle 21 to solar cycle 24. This fact indeed confirms the current trend to a sequence of progressive  weaker solar cycles. It is also interesting to see in the wavelet map the increasing extension of the "gap" (in blue) near the minimum phases of the cycles. In particular, the deep minimum for SC 23-24 is well evidenced with a greater extension both in time and in frequency. The Rieger-type period, near 157 d, is well confirmed for SC21, but less evident for next SC22 and SC23. Further, it seems absent for current SC24. This seems a non-persistent periodicity related to only high levels of magnetic activity. Thus, the suggested correlation with the emergence of magnetic flux in the photosphere needs to be better explained, if we consider weaker cycles like SC24 where some  magnetic activity is always present at least locally. The same bahavior is confirmed by other authors. In particular, Bisoi et al. report that the Rieger periodicity is a "well known fundamental periodicity that we find only in the high-latitude fields in the northern hemisphere prior to 1996."  

Periodicities near 3.5 yr have been already reported by other authors for unsigned photospheric fluxes. Recently, Bisoi et al. reported, using a wavelet analysis with the Torrence-Compo algorithm, periods in the range 3-5 yr between 1985 and 1996 for the high latitude photospheric fields. Moreover, the authors report that there is a clear transition in both periodicities and power levels in both solar hemispheres around 1996 for both high-latitude and low-latitude fields. From our Fig.1 we also see a change in the structure of the wavelet map after 1996 or julian date j2450212.5, with a lower power and a variation of some periodicities, but for the integrated magnetic indices this feature is less prominent. Periodicities in the interval: 209-222 d have been already reported by other authors for unsigned photospheric fluxes. From Fig.1 we see that this periodicity is quite prominent for MPSI index only in the current SC24.

In Fig. 2 we show the results of the wavelet analysis applied to the second magnetic index: MWSI. As for Fig.1, the time interval is: Jan 19, 1970 - Jan 22, 2013 The upper panel shows the original time series in its natural units. Note the weaker scale intensity. The central panel shows the amplitudes of the wavelet local power spectrum in terms of a color contour map. Horizontal time axis corresponds to the axis of the upper time series and the vertical scale (log frequency values in red) axis is expressed in period (days d or years yr). Here the range analyzed is: 2.8 d - 40 yr.
The right panel shows the global wavelet spectrum (an averaged and weighted Fourier spectrum) obtained by a time integration of the local map. The significance of the local power
map was tested using an adjustable red-noise autoregressive lag-1 background spectrum (Torrence and Compo, 1998).

\begin{figure}[h!]
\resizebox{\hsize}{!}{\includegraphics{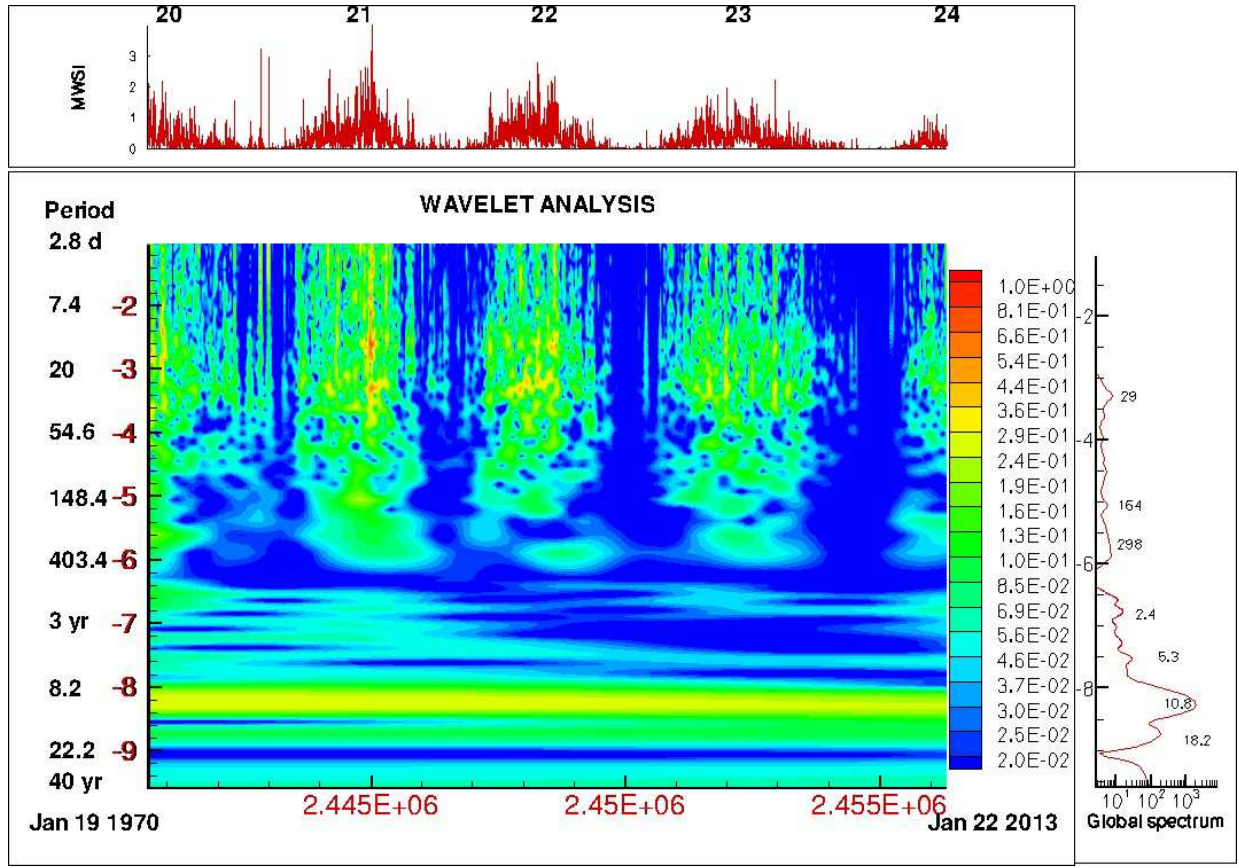}}
 \caption{Wavelet analysis of magnetic MWSI index. See text for details.}
 \label{fig2}
\end{figure}

The main periodicities detected are in decreasing order:

1) 18.2 yr 2) 10.8 yr; 3) 2.4 yr
5) 298 d; 6) 164 d; 7) 29 d

as we can see in the global power spectrum.

The MWSI index confirms the main features detected for MPSI index with some little differences. Here the reduction of power after SC22 is more evident and the intermediate periodicities are more difficult to point out. The periodicity of 2.4 yr is here evident only in the current SC24. Bisoi et al. report this "new" period using Fourier method for both north high and low latitude fields before 1996. Similarly, the 298 d periodicity is near the 294 d periodicity detected by Bisoi et al. for only north high latitude fields before 1996. On the other hand, Fig.2 shows that this periodicity is always present near the maximum phases for SC21-SC24.

Both the wavelet analyses of integrated magnetic indices MPSI and MWSI show that it is very useful to study the time-frequency behavior of local or global magnetic solar activity in order to better point out the changes or transitions that can drive the behavior of future solar magnetic fields and further these can give us a powerful tool to better interpret some "anomalous" behaviors observed, such as the recent deep minimum phase between SC23 and SC24 and the current decline in solar magnetic fields.

\section{Conclusions}

In this study we performed a wavelet analysis of integrated magnetic solar indices provided by the 150-Foot Solar Tower of the Mt. Wilson Observatory.  This analysis is a further contribution to detect the changes in the multiscale quasiperiodic variations in the integrated magnetic solar activity with a comparison between past solar cycles and the current one, which is one of the weaker recorded in the past 100 years. We confirme the current observed decline in solar magnetic fields reported by other authors and, more generally, a clear change in the structure of the wavelet map after 1996 with a lower power and a clear variation of some periodicities, that can be responsible for a state transition occurred in the magnetic mechanisms that drive the solar activity. Of course, we need both further data and detailed analyses in order to obtain a more clear and complete picture of the medium/long-term behavior of solar magnetic activity and its influences on space-weather conditions.

\section{Acknowledgements}

This study includes data from the synoptic program at the 150-Foot Solar Tower of the Mt. Wilson Observatory. The Mt. Wilson 150-Foot Solar Tower is operated by UCLA, with funding from NASA, ONR and NSF, under agreement with the Mt. Wilson Institute.

\section{References}

Bisoi, S.K., Janardhan, P. Chakrabarty, D., Ananthakrishnan, S. and Divekar, A., Changes in quasi-periodic variations of solar photospheric fields: precursor to the deep solar minimum in
the cycle 23?, 2013, arXiv:1304.8012v1 [astro-ph.SR]

Daubechies, I., Ten lectures on wavelets, 1992, SIAM, Philadelphia.

Grossmann, A., \& Morlet, J., Decomposition of Hardy functions into square integrable wavelets of constant shape, 1984, SIAM Jour. Math. Anal., 15, 723-736.

Lawrence, J.K., Cadavid, A.C., \& Ruzmaikin, A.A., Turbulent and chaotic dynamics underlying solar magnetic variability, 1995, ApJ, 455, 366.

Mallat, St$\acute{e}$fane, a Wavelet tour of signal processing, 1998, 2nd edition, Academic Press, San Diego, California.

Ochadlick, A.R., Kritikos, H.N., \& Giegengack, R., Variations in the period of the sunspot cycle, 1993, Geophys. Res. Lett., 1471-1474

Oliver, R., Ballester, J.L., \& Baudin, F., Emergence of magnetic flux on the Sun as the cause of a 158-day periodicity in sunspot areas, 1998, Nature, 394, 552-553.

Penn, M.J. \& Livingston, W. 2006, ApJ (Letters), 649, L45

Penn, M.J. \& Livingston, W., Long-term Evolution of Sunspot Magnetic Fields, 2010, arXiv:1009.0784v1 [astro-ph.SR]
 
Sello, S., Wavelet entropy as a measure of solar cycle complexity, 2000, A\&A, 363, 311-315.

Sello, S. Wavelet entropy and the multi-peaked structure of solar cycle maximum, New Astronomy, Volume 8, Number 2, February 2003, pp. 105-117(13).

Soon, W., Frick, P., Baliunas, S., Lifetime of surface features and stellar rotation: a wavelet time-frequency approach, 1999, ApJ, 510, 2, 135-138

The 150-Foot Solar Tower Current Selected Data: $http://obs.astro.ucla.edu/150_data.html$

Torrence, C., \& Compo, G.P., A practical guide to wavelet analysis, 1998, Bull. Am. Meteor. Soc., 79, 1, 61-78.

\end{document}